\documentclass[letterpaper]{IEEEtran}
\IEEEoverridecommandlockouts
\makeatletter
\def\markboth#1#2{\def\leftmark{\@IEEEcompsoconly{\sffamily}\MakeUppercase{\protect#1}}%
\def\rightmark{\@IEEEcompsoconly{\sffamily}\MakeUppercase{\protect#2}}}
\makeatother

\usepackage{kantlipsum}
\usepackage{setspace}

\usepackage[english]{babel}
\usepackage{xcolor}
\usepackage{lipsum}
\usepackage{caption}
\usepackage{cite}
\usepackage[pdftex]{graphicx}
\usepackage{subcaption}
\usepackage{amsmath}
\usepackage{booktabs}
\usepackage{colortbl}
\usepackage{amsfonts}
\usepackage{amssymb}
\usepackage{amsthm}
\usepackage{array}
\usepackage{verbatim}
\usepackage{listings}
\usepackage{algorithm}
\usepackage{algpseudocode}
\usepackage{url}
\usepackage{enumerate}
\usepackage{multirow}
\usepackage{xfrac}

\definecolor{LightBlue}{rgb}{0.5,0.5,1}
\definecolor{LightRed}{rgb}{1,0.5,0.5}
\definecolor{LightYellow}{rgb}{1,0.85,0}

\usepackage{epsfig}
\usepackage{epstopdf}
\usepackage{multicol}
\usepackage[font=footnotesize]{caption}

\usepackage{algorithm}

\makeatletter
\def\BState{\State\hskip-\ALG@thistlm}
\makeatother

\renewcommand{\arraystretch}{2}

\newcommand{\bi}{\begin{itemize}}
\newcommand{\ei}{\end{itemize}}
\newcommand{\be}{\begin{equation}}
\newcommand{\ee}{\end{equation}}

\def\beq{\begin{equation}}
\def\eeq{\end{equation}}
\def\beqa{\begin{eqnarray}}
\def\eeqa{\end{eqnarray}}
\def\beqan{\begin{eqnarray*}}
\def\eeqan{\end{eqnarray*}}

 \def\SINR{\mathsf{SINR}}
 \def\INR{\mathsf{INR}}
  \def\3G{\mathsf{3GPP}}  
    \def\ISO{\mathsf{ISO}}
    \def\H{\mathsf{HFSS}}

\title{Study of Realistic Antenna Patterns in 5G mmWave Cellular Scenarios}
\author{{{\textbf{Mattia Rebato\thanks{The authors would like to thank Dr. Danilo De Donno for useful discussions about this work.}}$^*$}, {\textbf{Laura Resteghini}}$^\dagger$, {\textbf{Christian Mazzucco}}$^\dagger$, {\textbf{Michele Zorzi}$^*$}}\\
\normalsize $^*$University of Padova, Italy and 
$^\dagger$HUAWEI Technologies, Milan, Italy \\
\small{$\{$\texttt{rebatoma}, \texttt{zorzi}$\}$\texttt{@dei.unipd.it}
} $\{$\texttt{laura.resteghini}, \texttt{christian.mazzucco}$\}$\texttt{@huawei.com}}

\begin{document}
\maketitle
\thispagestyle{empty}
\begin{abstract}

Large antenna arrays and millimeter-wave (mmWave) frequencies have been attracting growing attention as possible candidates to meet the high requirements of future 5G mobile networks.
In view of the large path loss attenuation in these bands, beamforming techniques that create a beam in the direction of the user equipment are essential to perform the transmission.
For this purpose, in this paper, we aim at characterizing realistic antenna radiation patterns, motivated by the need to properly capture mmWave propagation behaviors and understand the achievable performance in 5G cellular scenarios.
In particular, we highlight how the performance changes with the radiation pattern used.
Consequently, we conclude that it is crucial to use an accurate and realistic~radiation model for proper performance assessment and system~dimensioning. 
\vspace{-0.4cm}
\end{abstract}

\bigskip
\begin{picture}(0,0)(0,-345)
\centering
\put(0,0){
\put(0,0){ M. Rebato, L. Resteghini, C. Mazzucco, and M. Zorzi, ``Study of realistic antenna patterns in 5G mmwave cellular} 
\put(-28,-10){scenarios", in IEEE ICC Communications QoS, Reliability, and Modeling Symposium (ICC’18 CQRM), Kansas City, USA, May 2018.}}
\end{picture}

\vspace{-0.3cm}
\section{Introduction}
\label{introduction}

Due to high path loss attenuations, multiple-input multiple-output (MIMO) systems with beamforming techniques are essential to ensure an acceptable range of communication in millimeter-wave (mmWave) networks~\cite{rangan14}.
In particular, the use of antenna arrays for future mobile scenarios is fundamental in order to create a beam in the direction of the user equipment (UE), thus increasing the gain of the transmission. 
Among the possible antenna array designs, the most suitable approach is the use of uniform planar arrays (UPA) where the antenna elements are evenly spaced on a two-dimensional plane and a 3D beam can be synthesized~\cite{5a_book}. 

In~order~to~precisely evaluate 5G mmWave cellular~scenarios,~it~is~important to consider realistic and accurate radiation models. 
Related works in the literature characterize~the antenna array either over-simplifying its gain with piece-wise functions, or modeling it as an array of isotropic transmitting sources.
At high frequencies~(e.g.,~mmWave bands), where~high~attenuations~are~present, quantifying the actual~antenna gain obtained due to the radiation pattern~is fundamental in order to precisely evaluate any mmWave~system. 
\begin{figure}[t!]
\centering
\includegraphics[width=1\columnwidth]{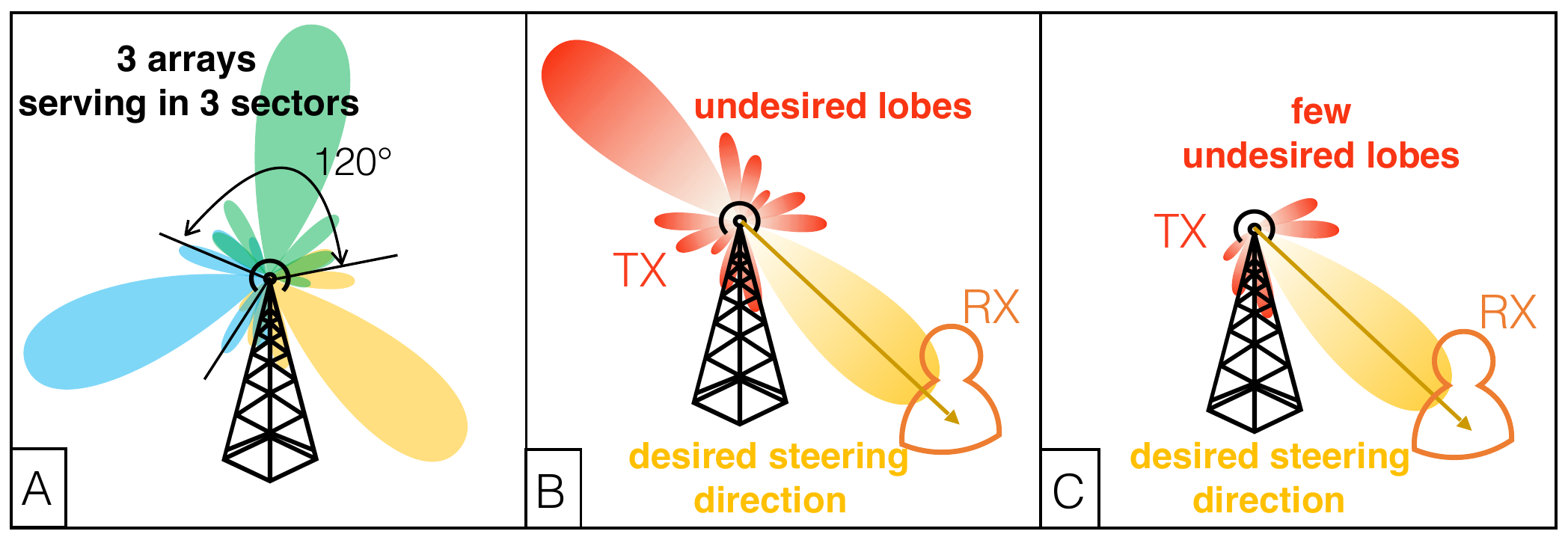}
\caption{Illustration of the different antenna configurations. \textbf{Scheme A}: Configuration with three arrays serving three sectors with the single-element $\3G$ antenna radiation pattern~\cite{mmwave_3gpp_channel}. \textbf{Scheme B}: Antenna radiation pattern using beamforming with \emph{isotropic} radiation elements. \textbf{Scheme C}: Antenna radiation pattern using beamforming with $\3G$ radiation elements.}
\vspace{-0.5cm}
\label{global_view}
\end{figure}
The radiation model proposed by the $\3G$ in~\cite{mmwave_3gpp_channel} can be~used to address this issue.
This model precisely simulates the~radiation pattern of a patch antenna element assuming large attenuation for lobes in the opposite plane of transmission.
In addition, as illustrated in Scheme A of Fig.~\ref{global_view}, the $\3G$ specifications suggest modeling each base station (BS) site with three~sectors, thus three arrays placed with central angles shifted by 120$^\circ$~each.  
The configuration with multiple sectors already~used in traditional 4G LTE systems seems to be~appropriate also for future cellular scenarios, and permits a~better control in the design of both desired and undesired lobes~(e.g., using arrays of patches that perform the steering within the interval $[-60^\circ,+60^\circ]$).

In this paper, motivated by the need to properly capture mmWave propagation behaviors and understand the achievable performance (e.g., capacity or interference studies) in 5G cellular scenarios, we aim at accurately characterizing the antenna radiation pattern. 
In addition to a realistic antenna pattern, we also incorporate mmWave channel characteristics as derived from a measurement-based mmWave channel model operating at 28~GHz provided by the New York University (NYU) Wireless Group.
This model is described in~\cite{akdeniz14,samimi15}, and was adopted in our previous works~\cite{rebato16,rebato16_interference}.

For tractability of analysis or ease of computation, most of the works in the literature approximate the actual beamforming patterns by a sector model.
An example of this approximation can be found in~\cite{bai15}, where a piece-wise beamforming gain function is used to characterize key features of an antenna pattern such as directivity gain, half-power beamwidth, and front-back ratio.
This model is too generic and cannot be compared with a realistic pattern because design parameters like gain, beamwidth and front-back ratio change according to the steering direction.
It is therefore challenging to compare this model with a realistic pattern since it uses fixed parameters for all the steerable directions.

A more precise antenna pattern can be obtained combining together the array factor expression, which provides information on the directivity equation of an antenna array, with the single element radiation pattern~\cite{tse_book}.
In~\cite{rebato16}, the array directivity equation with \emph{isotropic} antenna elements is applied.
As a result, each transmitting source (i.e., antenna element) radiates equal power in all directions, while a detailed lobe-shaped radiation pattern is used to capture the beamforming gain.
When considering beamforming with isotropic transmitting sources, undesired lobes are generated as shown in Scheme B of Fig.~\ref{global_view}.
Since BSs in real cellular systems do not transmit omnidirectionally but, instead, in sectors, it appears essential to consider antenna radiation models that avoid the generation of undesired lobes.
In this respect, a more realistic antenna pattern (e.g., the one shown in Scheme C of Fig.~\ref{global_view}) should be used when modeling UPA radiation.

Our study leads to the following observations. First, we highlight how the radiation pattern significantly influences the performance evaluation of cellular scenarios.
Results obtained with a simplified pattern appear to be significantly different from those obtained using realistic radiation models. 
Specifically, we try to quantify the interference perceived by a generic user in a way to identify the working regime (e.g., noise or interference limited) of mmWave networks. 
Second, we compare different realistic antenna patterns in order to evaluate how the design choices influence the entire network performance.   
Finally, we note that a proper estimation of the element radiation pattern should consider also the error introduced by the use of analog phase shifters.
In fact, although the beam steering performance depends on the number of bits used to quantize the phase shifter, the current state of silicon technologies makes the design of high-resolution mmWave phase shifters costly or even impractical~\cite{5a_book}.
To this end, we analyze the performance loss when considering an error in the synthesis of transmitter and receiver beams. 

\vspace{-0.3cm}
\section{Antenna array radiation patterns}
\label{antenna_array_radiation_model}

In this section we describe how to compute the realistic antenna array patterns considered in this study and report expressions for the diverse field factors.
The radiation pattern of the entire array (called array radiation pattern) is obtained by the superposition of its array factor, which provides information on the directivity equation of an antenna array, and the element radiation pattern.
This last term takes into account how power is radiated by each single antenna element.

We precisely describe in the following sub-sections how to obtain the pattern expression for the different antenna radiation models used in this comparison.

\paragraph*{\textbf{A. Element radiation pattern}}
The element radiation pattern indicated as $A_E$ and expressed in decibel (dB) is a term used to describe how the power of a single antenna element is radiated in all directions, and thus is defined for any pair of vertical and horizontal angles $(\theta,\phi)$.  
This parameter is extremely important in scenarios where directional transmission is used because it allows to precisely understand where the antenna transmits or receives~power.

For the purpose of evaluating the differences between the realistic patterns and those already available in the literature, we compare three different antenna radiation configurations.

\subsubsection{$\ISO$}
The \emph{isotropic} radiation pattern used in the literature is achieved using a single array of isotropic transmitting sources.
Hence, each element of the array redistributes equally the transmitted power in all directions, and beamforming is obtained considering only the array factor.
We use this radiation pattern for comparison, knowing that it is the least realistic.
It assumes that any antenna element irradiates an equal amount of power in each direction, thus: $A_E^{(\ISO)}=0$~dB$,\,\forall \theta \in [0,\pi],\forall \phi \in [-\pi,\pi]$.    
\begin{figure*}[t!]
        \centering
        \begin{subfigure}[b]{0.32\textwidth}
            \includegraphics[width=\textwidth]{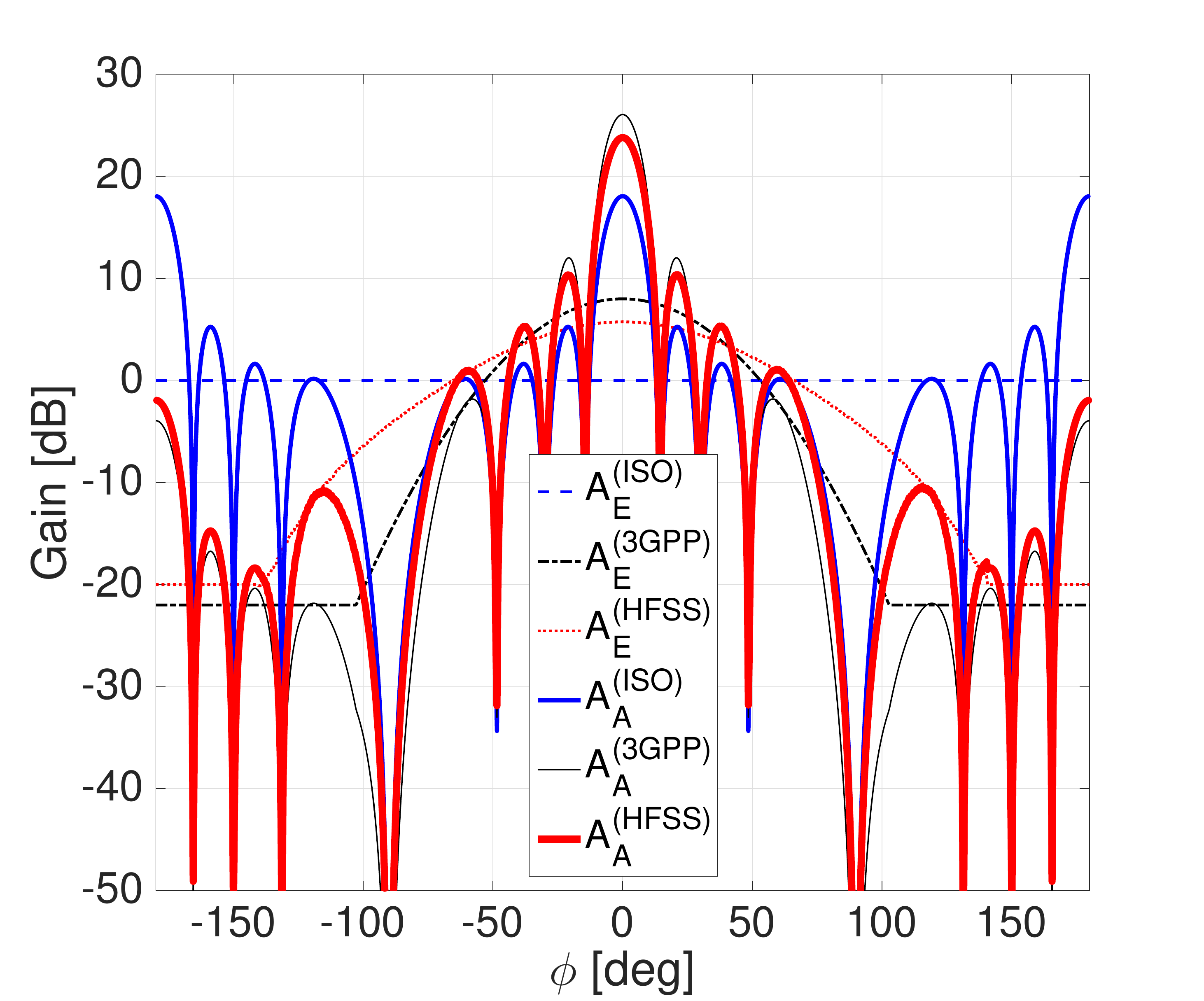}
            \caption{Configuration with $\phi_s = 0^\circ$.}
            \label{steering_0}
        \end{subfigure}
        ~ 
        \begin{subfigure}[b]{0.32\textwidth}
            \includegraphics[width=\textwidth]{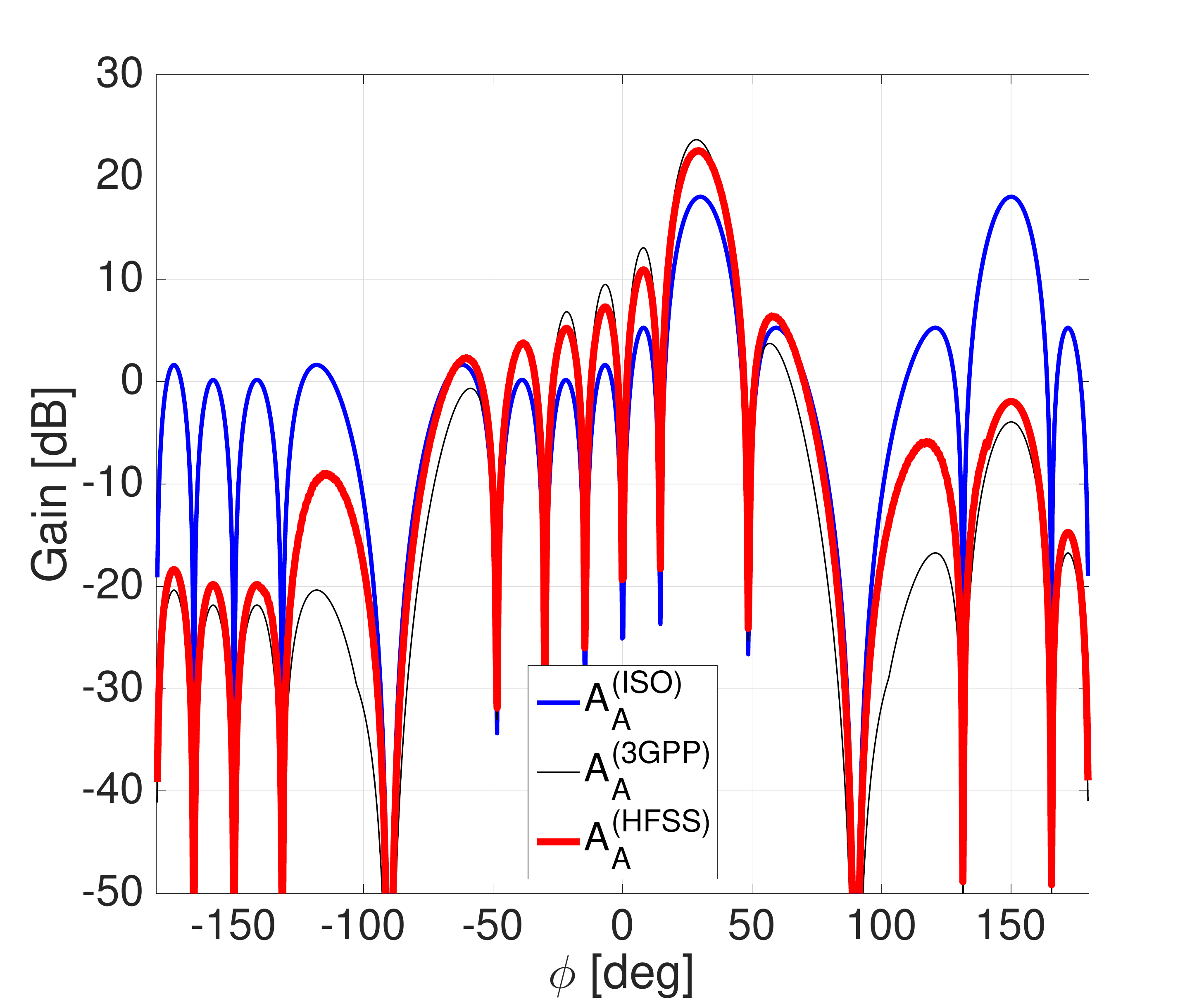}
            \caption{Configuration with $\phi_s = 30^\circ$.}
            \label{steering_30}
        \end{subfigure}
                ~ 
        \begin{subfigure}[b]{0.32\textwidth}
            \includegraphics[width=\textwidth]{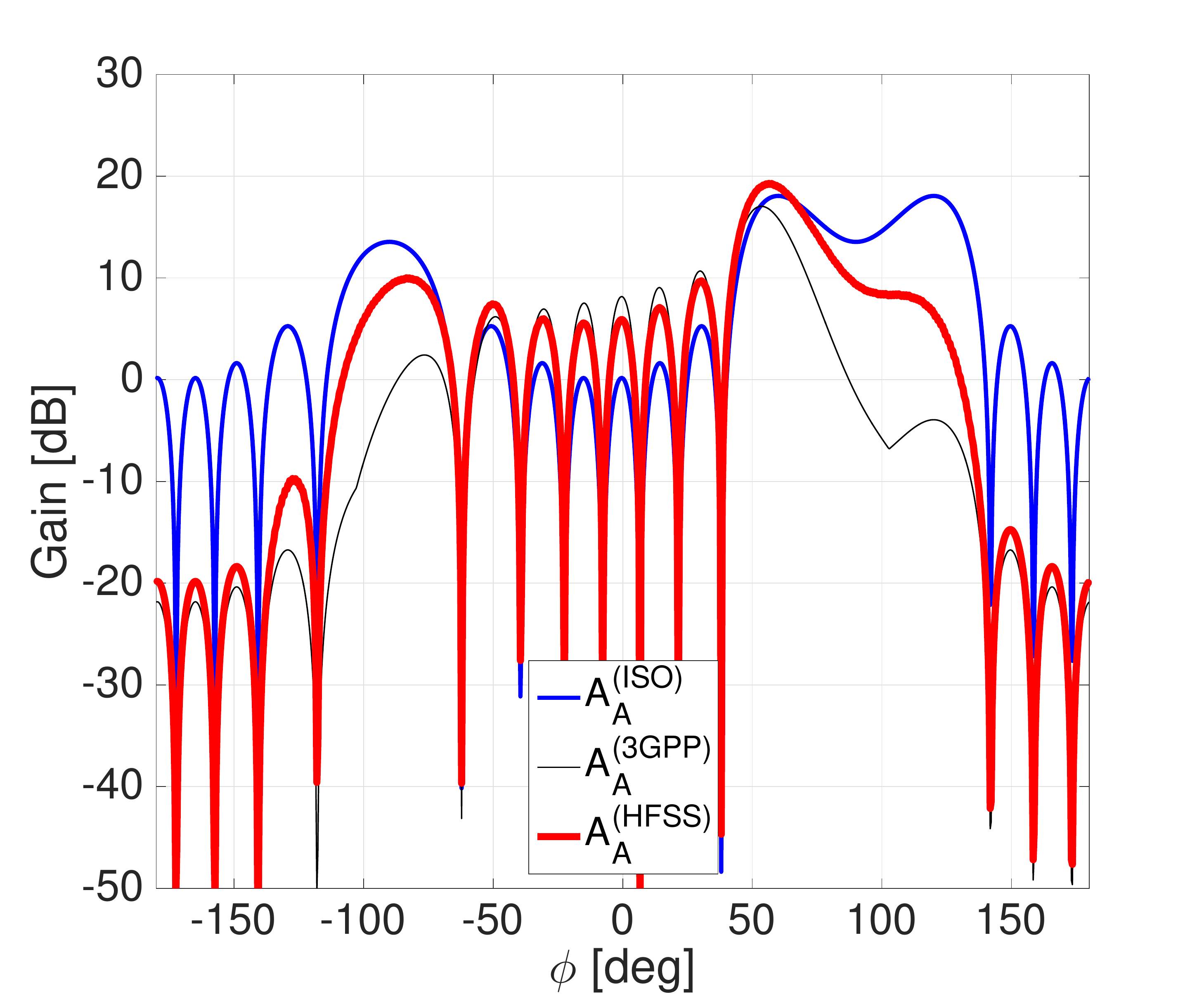}
            \caption{Configuration with $\phi_s = 60^\circ$.}
            \label{steering_60}
        \end{subfigure}
        \caption{Representation of the array radiation pattern $A_A(\theta,\phi)$ in relation with the element radiation pattern $A_E(\theta,\phi)$ varying horizontal angle $\phi$, while vertical angles $\theta$ and $\theta_s$ are kept fixed to 90$^\circ$. Examples obtained using a UPA with 64 antenna elements and performing the steering in different directions.}
        \label{fig_array_radiation_pattern}
    \end{figure*}
    
\subsubsection{$\3G$}
The $\3G$ model is realized following the specifications in~\cite{mmwave_3gpp_channel,antenna_3gpp,3d_channel_3gpp}.
First, differently from the $\ISO$ configuration, it implies the use of three sectors, thus three arrays, placed as in traditional mobile networks.
Second, the single element radiation pattern presents a high directivity with maximum gain in the main-lobe direction of about 8~dBi. 
The $\3G$ $A_E$ of each single antenna element is composed of horizontal and vertical radiation patterns.
Specifically, this last pattern $A_{E,V}(\theta)$ is obtained as
\begin{equation}
A_{E,V}(\theta) = - \min \left\{ 12 \left( \frac{\theta - 90}{\theta_{3 \text{dB}}}\right)^2, SLA_V \right\},
\end{equation}
where $\theta_{3 \text{dB}} = 65^\circ$ is the vertical 3~dB beamwidth, and $ SLA_V = 30 \text{ dB}$ is the side-lobe level limit.
Similarly, the horizontal pattern is computed as 
\begin{equation}
A_{E,H}(\phi) = - \min \left\{ 12 \left( \frac{\phi}{\phi_{3 \text{dB}}}\right)^2, A_m \right\},
\end{equation}
where $\phi_{3 \text{dB}} = 65^\circ$ is the horizontal 3~dB beamwidth, and $ A_m = 30$~dB is the front-back ratio.
Bringing together the previously computed vertical and horizontal patterns we can obtain the 3D antenna element gain for each pair of angles as
\begin{equation}
A_E^{(\3G)}(\theta,\phi) = G_{\max}- \min \left\{- \left[ A_{E,V}(\theta) + A_{E,H}(\phi) \right], A_m \right\},
\label{global_pattern_equation}
\end{equation}
where $G_{\max} = 8$~dBi is the maximum directional gain of the antenna element~\cite{mmwave_3gpp_channel}. 
The expression in~\eqref{global_pattern_equation} provides the dB gain experienced by a ray with angle pair $(\theta,\phi)$ due to the effect of the element radiation pattern.

\subsubsection{$\H$}
In addition to the $\3G$ pattern, we consider also a realistic pattern obtained reproducing a real patch antenna with a finite element simulator called High-Frequency Structural Simulator ($\H$)~\cite{web_hfss}.
This last model is the most realistic among the models studied, but is also the most computationally intensive.
As in the $\3G$ model, it considers a three-sector cell, while the element radiation pattern is modeled as a real patch antenna element working at 29.5~GHz with horizontal and vertical spacing and sizes equal to 0.55$\lambda$ and 0.77$\lambda$ respectively.
The $A_E$ of this last model is obtained by the $\H$ finite element simulator by setting the antenna element parameters (e.g., working frequency and size of the element), and exhibits a maximum gain of about 5.71~dBi.

\paragraph*{\textbf{B. Array radiation pattern}}
In order to study the pattern of a UPA, we must focus on the radiation of the entire array, i.e., taking into account the effect of all the elements. 

The relation between the array radiation pattern and a single element radiation pattern is defined, following~\cite{antenna_3gpp}, as
\begin{equation}
A_A^{(i)}(\theta,\phi) = A_E^{(i)}(\theta,\phi) + \mathsf{AF}(\theta,\phi).
\label{ralation_array_element}
\end{equation}
The expression in~\eqref{ralation_array_element} is valid for all three array radiation patterns, where the single-element term $A_E^{(i)}$ depends on the selected model.
More specifically, $i \in \{\ISO,\3G,\H\}$ corresponds to the element radiation functions described in the previous sub-section.
The relation in \eqref{ralation_array_element} considers the effect of the element radiation pattern in combination with the array factor $\mathsf{AF}(\theta,\phi)$ defined for an array of $n$ elements as
\begin{equation}
\mathsf{AF}(\theta,\phi) = 10 \log_{10}\left[ 1+ \rho \left( \left| \textbf{a} \cdot \textbf{w}^T \right|^2-1\right) \right]
\label{array_factor}
\end{equation}
where $\rho$ is the correlation coefficient, assumed equal to unity, $\textbf{a} \in \mathbb{C}^{n}$ is the amplitude vector and $\textbf{w} \in \mathbb{C}^{n}$ is the beamforming vector.
We assume to use an equal and fixed amplitude for all the antenna elements\footnote{
The array radiation pattern can be further refined adjusting the amplitude of each single element with a weighing factor.
This can provide both control of the side lobe levels and electrical steering.
For simplicity, in this work we consider an equal and fixed amplitude for each radiation element, and leave as future work the study of beamforming design optimizing the weight factor of each antenna element.}, thus $\textbf{a}$ is a constant normalized vector with all elements equal to $\tfrac{1}{\sqrt{n}}$.
The beamforming vector contains information about the main lobe steering direction $(\theta_s,\phi_s)$, and is obtained as
\begin{equation}
\begin{split}
\textbf{w} &= [w_{1,1},w_{1,2},  \dots,w_{m,m}], \text{ where } m = \sqrt{n},\\
w_{p,r} &= e^{ j 2 \pi \big( (p-1) \frac{\Delta_V}{\lambda} \Psi_p + (r-1) \frac{\Delta_H}{\lambda} \Psi_r \big) }, \\
&\begin{cases}
\Psi_p = \cos \left(\theta\right) - \cos \left(\theta_s\right) \\
\Psi_r = \sin \left(\theta\right) \sin \left(\phi\right) - \sin\left(\theta_s\right) \sin\left(\phi_s\right)
\end{cases}
\end{split}
\end{equation}
where $\Delta_V$ and $\Delta_H$ are the spacing distances between the vertical and horizontal elements of the array, respectively\footnote{Except for the $\H$ configuration, we always assume all elements to be evenly spaced on a two-dimensional plane, thus $\Delta_V = \Delta_H = \lambda / 2$.}.
A detailed explanation of the relation between array and element patterns can be found in~\cite{antenna_3gpp}.
We highlight that the pair of angles $(\theta,\phi)$ must not be confused with the steering pair $(\theta_s,\phi_s)$ where the main beam is steered due to beamforming.

For ease of computation, we are not considering mutual coupling\footnote{
Mutual coupling describes energy absorbed by one antenna's receiver when another nearby antenna is operating.
This effect is typically undesirable because the energy that should be radiated away is absorbed by a nearby antenna. Similarly, energy that could have been captured by one antenna is instead absorbed by a nearby antenna. Hence, mutual coupling reduces the antenna efficiency and performance of antennas in both the transmit and receive mode.
} effects in our comparison study. 

\paragraph*{\textbf{C. Field pattern}}
In order to integrate the array radiation pattern into the channel model, we must compute the field pattern~\cite{3d_channel_3gpp}, which comprises both vertical and horizontal polarization terms as
\begin{equation}
\begin{cases}
F_{\theta}^{(i)} (\theta,\phi) = \sqrt{A_A^{(i)}(\theta,\phi)}\cos (\zeta),\\
F_{\phi }^{(i)} (\theta,\phi) = \sqrt{A_A^{(i)}(\theta,\phi)}\sin (\zeta),
\end{cases}
\end{equation} 
respectively, where $\zeta$ is the polarization slant angle and $A_A^{(i)}(\theta,\phi)$ is the 3D antenna array gain pattern previously obtained in~\eqref{ralation_array_element}.
Note that, for simplicity, in this study we consider a purely vertically polarized antenna, thus $\zeta = 0$ and
\begin{equation}
\begin{cases}
F_{\theta}^{(i)} (\theta,\phi) = \sqrt{A_A^{(i)}(\theta,\phi)} = F^{(i)} (\theta,\phi),\\
F_{\phi}^{(i)} (\theta,\phi) = 0.
\end{cases}
\end{equation} 

In Fig.~\ref{fig_array_radiation_pattern} we report examples of the three array radiation patterns  for different steering directions.
The figures highlight two important aspects.
Firstly, they show the reduction of the undesired side lobes when considering beamforming with $\3G$ and $\H$ antenna models.
An accurate evaluation of the interference can be done only knowing the precise gain in all directions: unrealistic side lobes must be removed from the pattern while, at the same time, realistic lobes must be properly considered.  
Secondly, as we can see in Fig.~\ref{steering_60}, the maximum gain is affected by the scan loss\footnote{A problem associated with beam scanning is the beam distortion with the scan angle. Steering at the side of the array results in the spread of the beam shape and a consequent reduction in gain known as \emph{scan loss}.} when the main lobe is steered from the broadside direction: due to the directivity of the single antenna element radiation pattern, the main lobe obtained with beamforming has beamwidth and maximum gain that depend on the steering angle.
In fact, $\phi_s = 60^\circ$ is the maximum angle allowed with the 3-sector configuration\footnote{With three sectors each array can perform the steering within the interval $[-60^\circ,+60^\circ]$. } and in this particular case the main beam presents a gain that is lower if compared to the other smaller steering angles in Figs.~\ref{steering_0} and~\ref{steering_30}.
Moreover, this can also be observed in the element radiation pattern $A_E$ in Fig.~\ref{steering_0}, where the $\3G$ configuration exhibits an attenuation of 10~dB for $\phi = 60^\circ$ with respect to the central angle $\phi = 0^\circ$.
On the contrary, the $\H$ configuration presents a smaller attenuation of around 6~dB.
Indeed, we notice that the $\H$ element radiation pattern is less directive compared to the $\3G$ model, being the maximum gain of the single-element radiation pattern around 5.7~dBi for the former and 8~dBi for the latter.
Such a difference in directivity results in higher gain for the $\3G$ model when a central steering is performed, while on the contrary higher gain is exhibited by the $\H$ configuration when steered broadside (e.g., around 60$^\circ$).

\vspace{-0.4cm}
\section{System model}
\label{system_model}

\paragraph*{\textbf{A. Channel characterization}}
    We adopt the NYU channel model presented in~\cite{akdeniz14,samimi15}, which is derived from the WINNER~II model~\cite{winner2} and is based on real-world measurements at 28~GHz.
    According to the channel model, each wireless link comprises $K$ clusters, corresponding to macro-level scattering paths, in turn composed of $L_k$ subpaths. 

    Given a set of clusters and subpaths, each element of the channel matrix $\textbf{H}\in \mathbb{C}^{n_{\text{TX}}\times n_{\text{RX}}}$, which characterizes a communication link, is represented as
    \begin{equation}
    h_{r,t}= \sum_{k=1}^{K}\sum_{l=1}^{L_k}g_{kl} F_{r}\left(\Omega^{r}_{kl}\right)u_{r}\left(\Omega^{r}_{kl}\right) F_{t}\left(\Omega^{t}_{kl}\right)u^*_{t}\left(\Omega^{t}_{kl}\right),
    \label{channel_matrix}
    \end{equation}
    where $t$ and $r$ are the indices of the $t$-th and $r$-th elements of the transmitter and receiver array respectively, $g_{kl}$ is the small-scale fading gain of subpath $l$ in cluster $k$, $F_r$ and $F_t$ are the receiver and transmitter field patterns previously computed, and $u_{r}(\cdot)$ and $u_{t}(\cdot)$ indicate the 3D spatial signature element of the receiver and transmitter, respectively.
    Moreover, $\Omega^{r}_{kl} = \left(\theta^{r}_{kl},\phi^{r}_{kl}\right)$ are the angular spread of vertical and horizontal angles of arrival and   $\Omega^{t}_{kl} = \left(\theta^{t}_{kl},\phi^{t}_{kl}\right)$ are the angular spread of vertical and horizontal angles of departure, both for subpath $l$ in cluster $k$~\cite{akdeniz14}.
    Note that even if we consider channels with UPA antennas, in our simulations we neglect for ease of computation the vertical signatures by setting their angles $\theta_{(\cdot)}$ equal to $\pi/2$.
As defined herein, the channel matrix $\textbf{H}$ contains information on the channel conditions along with beamforming and antenna radiation pattern. 

    Then, the small-scale fading gain $g_{kl}$ is given as follows
    \begin{align}
    g_{kl}=\sqrt{P_{kl}}e^{-j2\pi \tau _{kl}f},
    \label{scale_fading}
    \end{align}
where $P_{kl}$ denotes the power gain of subpath $l$ in cluster $k$, $\tau _{kl}$ is the delay spread induced by different subpath distances, and $f$ indicates the carrier frequency. Specific parameters are provided in Table I.

    
    Consider a directional beamforming where the main lobe center of a BS's transmit beam points at its associated UE, while the main lobe center of a UE's receive beam aims at the serving BS.
    Unless stated otherwise, we assume that both beams can be steered in any direction.
    Therefore, in each sector, we can generate a beamforming vector $\textbf{w}$ for any possible angle in the interval $[-60^\circ,+60^\circ]$.\footnote{We recall that, among the different configuration, only the $\ISO$ model uses a single sector.
    Therefore, the $\ISO$ beam can be generated in any direction.}
    At a typical UE $j$, the aligned gain $G_{ij}$ (considering also the channel) is its beamforming gain towards the serving BS $i$.
    With a slight abuse of notation, we represent it as
    \begin{equation}
    G_{ij} = \left| \sum_{r=1}^{n_{\text{RX}}} \sum_{t=1}^{n_{\text{TX}} } h_{r,t} \right|^2.
    \label{bf_gain_1}
    \end{equation}

\begin{table}
    \centering
    \caption{List of notations and channel parameters.}
    \small
    \renewcommand{\arraystretch}{1}
    \begin{tabular}{r |l }
    \toprule
    \hspace{-10pt}\bf{Notation} &\hspace{-5pt} \textbf{Meaning}\\
    \hline 
         \hspace{-10pt} $A_E$ & Element radiation pattern\\
                  \hspace{-10pt} $\mathsf{AF}$ & Array factor\\
          \hspace{-10pt} $A_A$ & Array radiation pattern\\
           \hspace{-10pt} $F$ & Field pattern\\
    \hspace{-10pt} $\ell(r)$ & Path loss at distance $r$ in LoS/NLoS/out states\\
    \hspace{-10pt} $n_\text{TX},\; n_\text{RX}$ & \# antennas of a BS and a UE\\
    \hspace{-10pt} $K$ & \# clusters $\sim \max\{\textsf{Poiss}(1.8),1\}$ \\
    \hspace{-10pt} $L_k$ & \# subpaths in the $k$-th cluster $\sim \textsf{DiscreteUni}[1,10]$ \\
    \hspace{-10pt} $\phi_{kl}^{r}$, $\phi_{kl}^{t}$ & Angular spread of subpath $l$ in cluster $k$~\cite{akdeniz14}: \\
    \hspace{-10pt} & $\phi_{k}^{(\cdot)}\hspace{-3pt}\sim \textsf{Uni}[0,2\pi]$, $s_{kl}\sim \max\{\textsf{Exp}(0.178),0.0122\},$\\
    \hspace{-10pt} & $\phi_{kl}^{(\cdot)}=\phi_{k}^{(\cdot)} + (-1)^l s_{kl}/2$\\
    \hspace{-10pt} $P_{kl}$ & Power gain of subpath $l$ in cluster $k$~\cite{samimi15}:\\
    \hspace{-10pt} & $U_k \sim \textsf{Uni}[0,1]$, $Z_k \sim \mathcal{N}(0,4^2)$, $V_{kl} \sim \textsf{Uni}[0,0.6]$, \\
        \hspace{-10pt} & $\tau_{kl}=2.8, P_{kl}=\tfrac{P_{kl}^\prime}{\sum P_{kl}^\prime}$, $P_{kl}^\prime = \frac{U_k^{\tau_{kl}-1}10^{-0.1 Z_k+V_{kl}}}{L_k}$
        \\
    \bottomrule
    \end{tabular}
    \label{Table:Notations}
     \vspace{-0.3cm}
    \end{table}

\paragraph*{\textbf{B. $\INR$ and $\SINR$ definitions}}
Thanks to a detailed channel and antenna characterization, we can compute the signal-to-interference-plus-noise ratio ($\SINR$) between transmitter $i$ and receiver $j$ as
\begin{equation}
\SINR_{ij} = \frac{\frac{P_{\text{TX}}}{\ell_{ij}}G_{ij}}{\sum_{y \neq i} \frac{P_{\text{TX}}}{\ell_{yj}}G_{yj} + W \times N_0},
\label{equation_sinr}
\end{equation}
where $y$ represents the $y$-th interfering link, $W$ is the total bandwidth, $N_0$ is the thermal noise, $P_{\text{TX}}$ is the transmitted power, $G$ is the beamforming gain, and $\ell$ is the path loss.
This last quantity is modeled with three states, as reported in~\cite{akdeniz14}: LoS, NLoS and outage, as a function of the distance $d$ between transmitter and receiver. 
Furthermore, in our simulations, we associate each UE to the BS that provides the smallest path loss (maximum average received power). 

Similarly to the $\SINR$, we can compute the interference-to-noise ratio ($\INR$), which is defined following~\cite{rebato16_interference} as 
\begin{equation}
\INR_{ij} = \frac{\sum_{y \neq i} \frac{P_{\text{TX}}}{\ell_{yj}}G_{yj}}{W \times N_0}.
\label{inr_equation}
\end{equation}

\vspace{-0.4cm}
\section{Simulations and results}
\label{simulations_and_results}

In this section, we report simulation results to show the achievable performance of the three different antenna configurations.
We consider scenarios using UPAs with 64 antenna elements (8$\times$8) at the BS, while at the receiver side we model UPAs with 16 elements (4$\times$4).
In our simulation campaign, we have adopted a 7~dB noise figure, a 500~MHz total bandwidth, and a transmit power $P_{\text{TX}}=30$~dBm, which are in line with the specifications envisioned for downlink transmission in 5G mmWave mobile networks~\cite{rangan14}.

The first result we report is a study on the level of interference experienced by a generic user in a 5G cellular network.
In order to quantify the amount of interference, we compute the $\INR$ as in~\eqref{inr_equation} for a large number of independent simulations and then derive its empirical cumulative distribution function (ECDF).
With the ECDF we identify the point that satisfies the condition $\INR = 1$ as the transitional point that determines the shift from a noise-limited to an interference-limited regime.
In this manner, we obtain the user noise-limited probability $\mathsf{P}_{\mathsf{NL}}$, defined as the probability for a generic user to have the noise power bigger than the interference power.
This metric allows to better understand the behavior of a network for different~BS~densities.

\begin{figure}[t!]
\centering
\includegraphics[width=0.73\columnwidth]{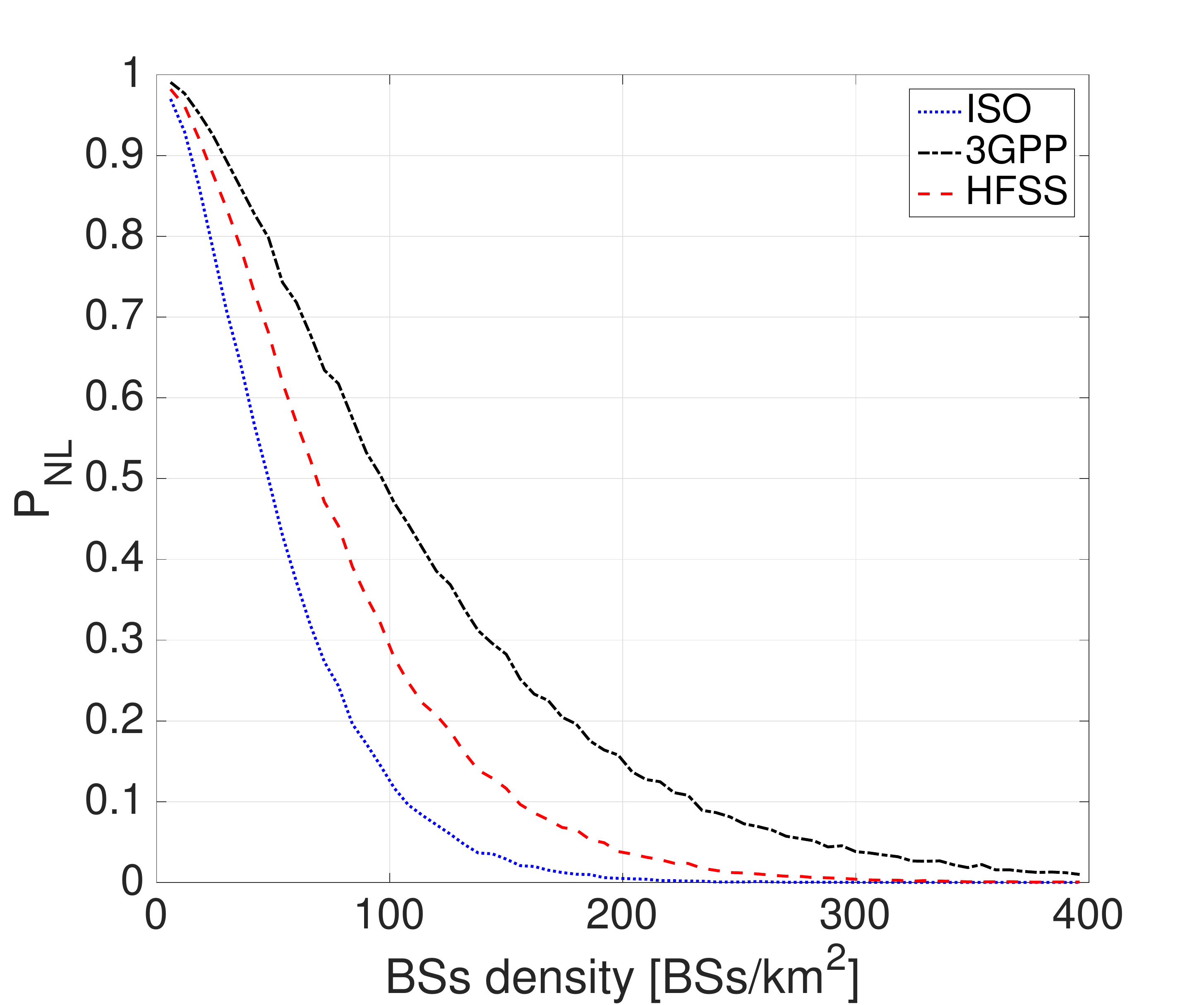}
\caption{Noise-limited probability $\mathsf{P}_{\mathsf{NL}}$ varying the BS density in the three different configurations.}
\vspace{-0.4cm}
\label{p_nl}
\end{figure}
\begin{figure*}[t!]
        \centering
        \begin{subfigure}[b]{0.32\textwidth}
            \includegraphics[width=\textwidth]{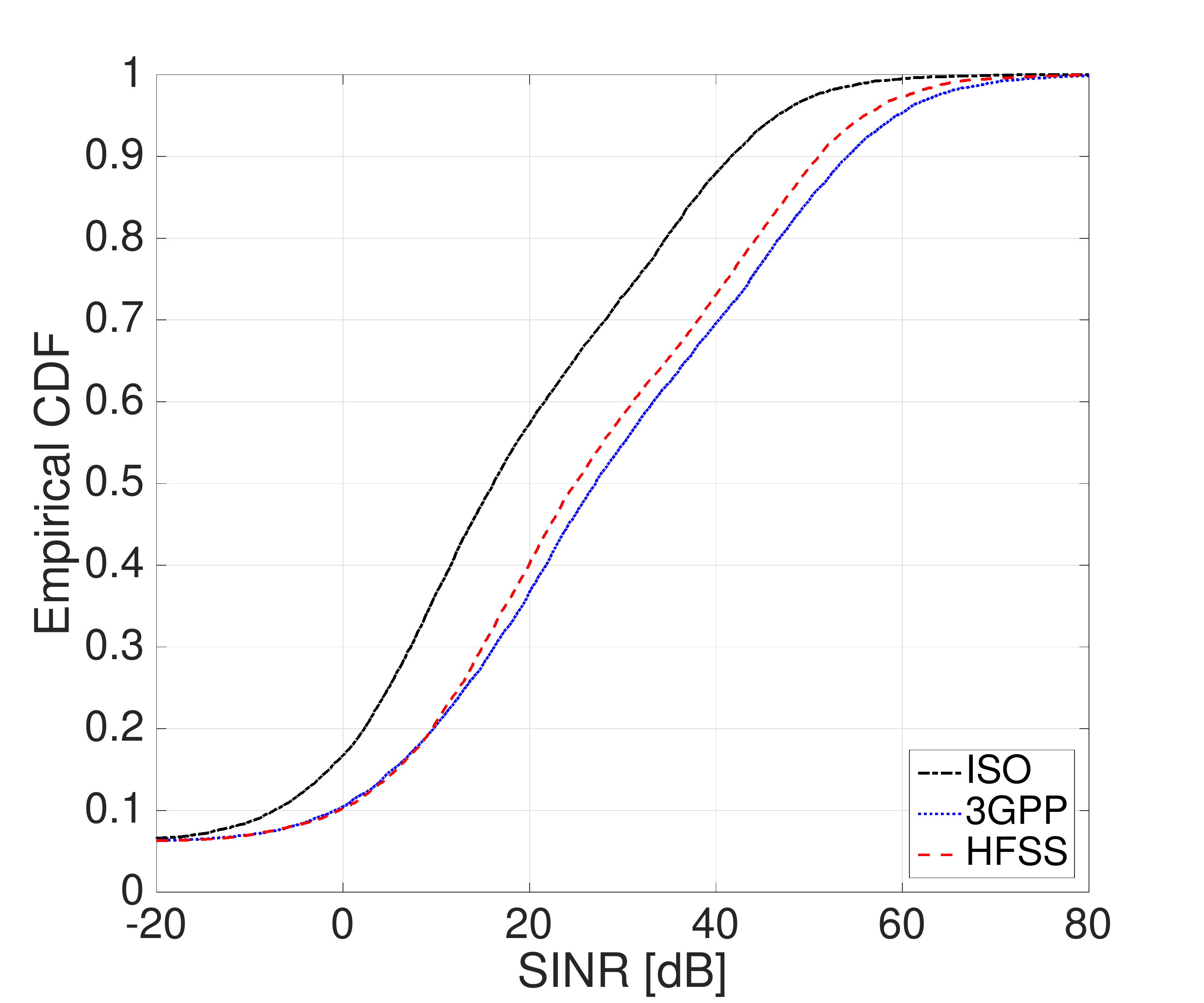}
            \caption{Example of varying pattern with density equal to 25~BSs/km$^2$.}
            \label{density_25}
        \end{subfigure}
        ~ 
        \begin{subfigure}[b]{0.32\textwidth}
            \includegraphics[width=\textwidth]{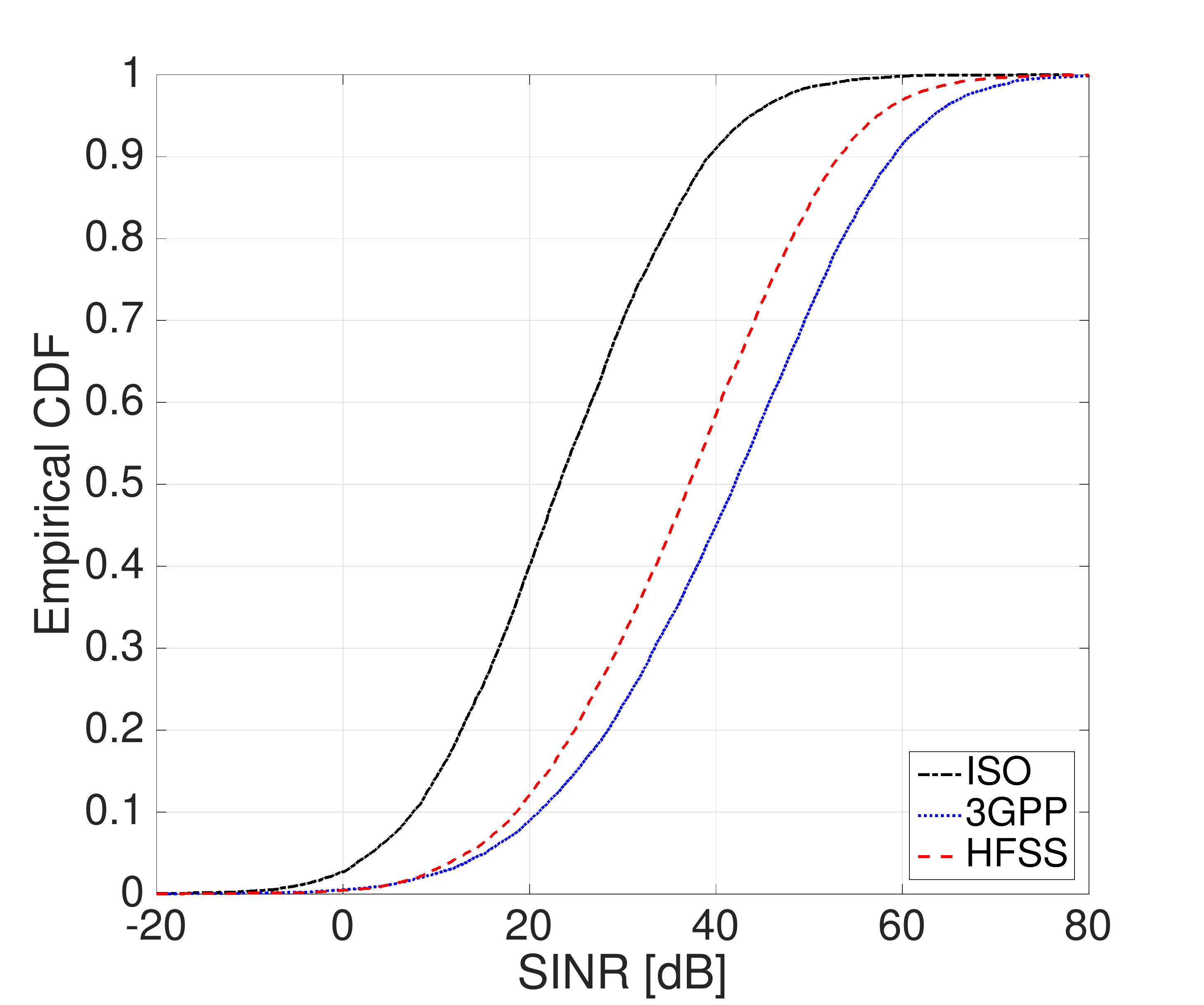}
            \caption{Example of varying pattern with density equal to 100~BSs/km$^2$.}
            \label{density_100}
        \end{subfigure}
                ~ 
        \begin{subfigure}[b]{0.32\textwidth}
            \includegraphics[width=\textwidth]{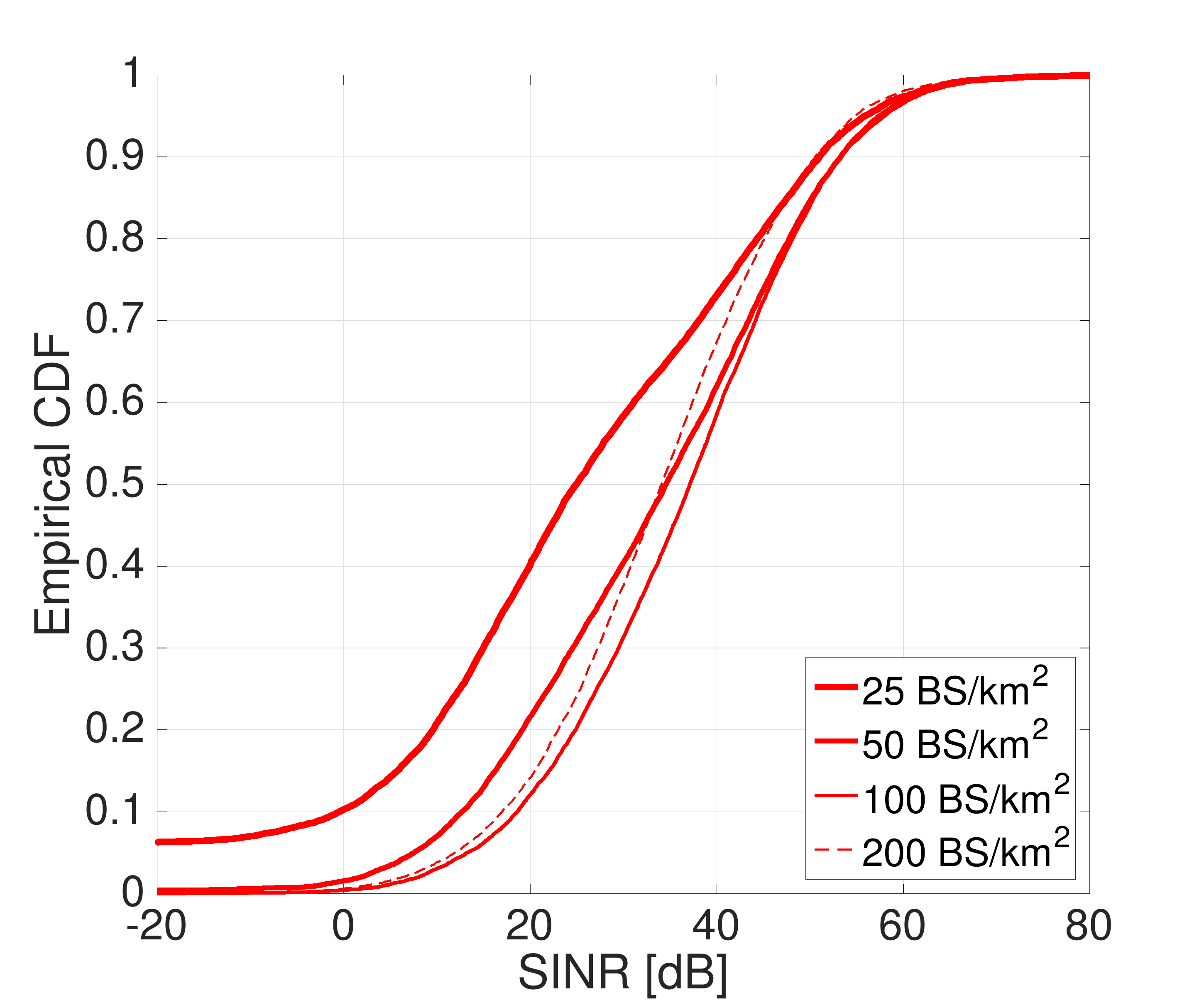}
            \caption{Example of varying BS density with $\H$ radiation pattern.}
            \label{density_hfss}
        \end{subfigure}
        \caption{Representation of Empirical CDF of the downlink $\SINR$, for different antenna configuration and varying the BS density.}
        \label{fig_sinr}
    \end{figure*}
In Fig.~\ref{p_nl} we show the evolution of $\mathsf{P}_{\mathsf{NL}}$ for the different radiation pattern configurations and varying the BS density of the network.
We can notice immediately that $\mathsf{P}_{\mathsf{NL}}$ is bigger in both $\3G$ and $\H$ with respect to the $\ISO$ configuration.
This is due to the attenuation of the interference in both $\3G$ and $\H$ configurations caused by the use of directive elements which attenuate the side lobes. 
Furthermore, the interference is larger in the $\ISO$ configuration due to the use of a single sector that operates in all directions. 
We recall that both $\3G$ and $\H$ configurations use three sectors (i.e., three arrays) that are orientated with a mechanical phase-shift of 120$^\circ$ each as in traditional cellular networks.
As we can see from Fig.~\ref{p_nl}, for densities around 100~BSs/km$^2$ and using an $\ISO$ antenna configuration, about 90\% of the users can be considered in an interference-limited regime.
This is not verified when adopting the $\H$ or $\3G$ antenna configurations, in fact, for the same density, only 70\% and 50\% of the users are in an interference-limited regime for the $\H$ and $\3G$ pattern, respectively.
More precisely, using the $\3G$ antenna model only half of the users are in an interference-limited scenario with a density of 100 BSs/km$^2$.
This corresponds to a dense scenario where the average cell radius is about 60 meters.

Differently from prior $\INR$ studies in~\cite{rebato16_interference,fischione18,rangan14} and~\cite{heath15}, where antenna patterns were modeled with simplified functions for tractability, here we can precisely evaluate the operating regime of mmWave networks as a function of the BS density.  
We also highlight how the interference results change with the adopted radiation pattern, which confirms the importance of using a precise radiation model when evaluating the network performance.

In Fig.~\ref{fig_sinr}, we show the ECDF of the downlink $\SINR$, for each antenna configuration, varying the BS density.
Firstly, as per the result in the interference study, also in this $\SINR$ evaluation we can see how the performance of a typical network changes with the different antenna patterns.
The different behavior can easily be seen starting from Fig.~\ref{density_25} where, with 25~BS/km$^2$, the median value of the $\ISO$ configuration is around 10~dB, while in the $\3G$ or $\H$ configuration it is above 20~dB. 
Moreover, with a small density of 25~BS/km$^2$, all curves show the presence of users in an \emph{outage} condition, which is revealed by the $\SINR$ ECDF not starting from zero. 
Increasing the BS density from Fig.~\ref{density_25} to Fig.~\ref{density_100} we can see how the $\SINR$ improves while maintaining different outcomes for the different pattern configurations.
Also, with the larger density, no users are in an \emph{outage} condition.

We can notice another interesting result passing from 100 to 200 BS/km$^2$ in Fig.~\ref{density_hfss}, where the larger amount of interference due to a larger number of transmitting sources results in decreased performance.
Contrary to the performance improvement for the smaller density case, with 200~BS/km$^2$~the reduced distance between UE and BS provides less improvement with respect to the growth of the interference.
A similar behavior has been observed also in the $\ISO$ and $\3G$~models.

\begin{table}
\centering
\renewcommand{\arraystretch}{0.98}
\caption{Summarizing table reporting the 5th ECDF percentile of the $\SINR$ for different configurations and densities. Values are expressed in decibel.}
\label{table_sinr}
\begin{tabular}{r|c|c|c}
\toprule
\multicolumn{1}{l|}{} & \multicolumn{3}{c}{BS/km$^2$} \\ \cline{2-4} 
\multicolumn{1}{l|}{} & 50       & 100      & 200      \\ \hline \hline
$\ISO$                & 0.00     & 3.02     & 1.74     \\ \hline
$\3G$                 & 8.11     & 15.31    & 14.03    \\ \hline
$\H$                & 7.54     & 13.22    & 11.94   \\
\bottomrule
\end{tabular}
\vspace{-0.5cm}
\end{table}
Table~\ref{table_sinr} reflects the results shown in Fig.~\ref{fig_sinr} and provides a comparison among all the different antenna setups and BS densities.
Specifically, we report the 5th percentile values of the user $\SINR$ in dB, which represents the performance of the worst users and is typically used to classify network performance. 
Note that $\SINR$ values for the density of 25 BS/km$^2$ are not reported in the table because more than 5\% of the users are in an \emph{outage} condition with such a small BS density. 
In general, our results show that different antenna models may provide quite different $\SINR$ performance. Also, we observe a performance gain passing from 50 to 100 BS/km$^2$.
Instead, as already discussed for Fig.~\ref{density_hfss}, in extremely dense networks (e.g., with 200~BS/km$^2$) the interference is really large and we observe $\SINR$ values that are smaller than in the case with 100~BS/km$^2$.
(This behavior is the same for all the radiation patterns used.)

\begin{figure}[t!]
        \centering
            \includegraphics[width=0.73\columnwidth]{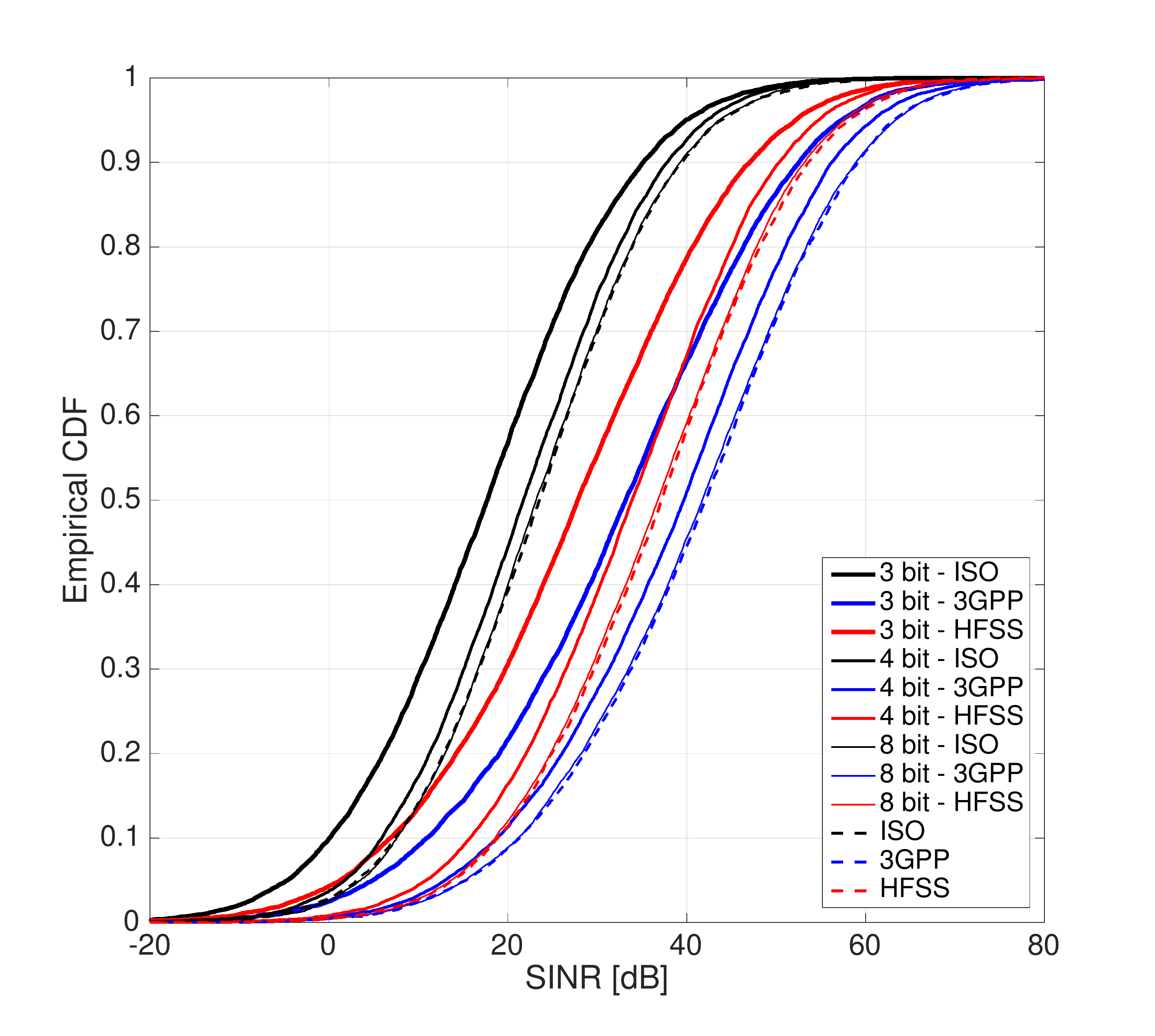}
            \caption{ECDF of the $\SINR$ for different phase shifter resolutions. Values of simulations with a BS density of 100 BSs/km$^2$.}
            \vspace{-0.3cm}
            \label{sinr_quantization_bit}
\end{figure}

A further study is presented in Fig.~\ref{sinr_quantization_bit}, where we report the ECDF of the $\SINR$ varying the phase shifter resolution in terms of the number of quantization bits.
In the figure, the dashed lines correspond to the upper bound cases in which we assume that both transmitter and receiver beams are perfectly aligned because beams can be steered in any direction (i.e., using an infinite number of quantization bits). 
As already mentioned, since an accurate study should consider also the error introduced in the synthesis of real beams, we report here how the $\SINR$ evolves passing from 3 to 8 quantization bits for each phase shifter.
In this way we can also evaluate the correctness of the results when errors in the beam synthesis are not considered.
   This result shows how, differently from the $\SINR$ changes for the diverse radiation patterns, quantization is less varying and disparities in the performance are limited.
Moreover, it can be observed how the performance deteriorates (i.e., $\SINR$ decreases) using a small number of bits, due to the reduced capability of the synthesized radiation pattern to align the beams. 
Instead, when 8 quantization bits are used, the achieved performance is very close to the upper bound, meaning that further increasing the phase shifter resolution does not improve the $\SINR$ performance.
Nevertheless, as the figure shows, the use of 4~bits represents a good trade-off between performance and phase shifter cost.   

\vspace{-0.2cm}
\section{Conclusion and future works}
\label{conclusion_and_future_works}

In this paper we have discussed the importance of precisely characterizing the antenna radiation pattern when studying 5G mmWave cellular scenarios. 
Our study led to the following observations.
First, we have highlighted how the radiation pattern influences the performance evaluation of cellular scenarios.
Specifically, as seen in the results, we have quantified the interference perceived by a generic user in a way to identify the working regime (e.g., noise or interference limited) of mmWave networks. 
Second, we have compared different realistic antenna patterns providing an understanding of how the design choices influence the entire network performance.   
Finally, we have studied how the performance behaves when considering an error in the alignment between transmitter and receiver beams.
Results show that good performance can be obtained using phase shifters with only 4~bits. 

We leave as a future work the optimization of array factor components such as the spacing of the elements, the amplitude and the phase vectors of each antenna element.
Without changes in the physical antenna elements, the array factor design can be adjusted on-the-fly, by tuning both amplitude and beamforming vectors for each user, in a way to improve the overall network performance.

\bibliographystyle{IEEEtran}
\bibliography{biblio}

\end{document}